\documentclass[epj]{svjour}
\usepackage{graphics}

\begin{document}
\title{ The contributions of $qqqq\bar{q}$ components to the axial charges of proton and its resonances}
\subtitle{\\}

\author{{S. G. Yuan}
  \inst{1,2,}\thanks{\email{yuanhadron@impcas.ac.cn}} \and
  {C. S. An}
\inst{3,}\thanks{\email{ancs@ihep.ac.cn}} \and {Jun He}
\inst{1,}\thanks{\email{junhe@impcas.ac.cn}} }

\institute{Institute of Modern Physics, Chinese Academy Sciences,
Lanzhou 730000, China \and Graduate University of Chinese Academy
Sciences, Beijing 100049, China \and Institute of High Energy
Physics, Chinese Academy Sciences, P.O.Box 918, Beijing 100049,
China}
\date{Received: date / Revised version: date}

\abstract{ In this paper we calculate the axial charges of the
proton and its resonances in the framework of the constituent quark
model, which is extended to include the $qqqq\bar{q}$ components. If
$20\%$ admixtures of the $qqqq\bar{q}$ components in the proton are
assumed, the theoretical value for the axial charge in our model is
in good agreement with the empirical value, which can not be well
reproduced in the traditional constituent quark model even though
the $SU(6) \bigotimes O(3)$ symmetry breaking or relativistic effect
is taken into account. We also predict an unity axial charge for
$N^{*}(1440)$ with $30\%$ $qqqq\bar{q}$ components constrained by
the strong and electromagnetic decays.
}

\maketitle

\section{Introduction}

\label{sec:1}

Axial charges of the nucleon and nucleon resonances are fundamental
quantities in QCD. They quantify spontaneous chiral symmetry
breaking in the low energy QCD. The proton axial charge is well
known from neutron beta decay experiment, $g_{A}$=$1.2670\pm 0.0030$
(in units of vector charge $g_{V}$) \cite{pdg}. The conventional
quark model in the $SU(6) \bigotimes O(3)$ symmetry scheme  predicts
that axial charge of the proton is exactly 5/3, which is about
$24\%$ larger than experimental value as given above. The diagonal
axial charges of nucleon resonances can not be measured
experimentally, but they can be calculated on the lattice QCD.
Recently the lattice results show that $N^{*}(1535)$ has a very
small axial charge and the axial charge of $N^{*}(1650)$ is about
0.55 \cite{ta}, which agree with constituent quark model (CQM)
values under the $SU(6) \bigotimes O(3)$, -1/9 and 5/9 ,
respectively. These values imply that the mixing between the two
negative states is small, which conflicts with the popular view that
the mixings are required by the data under very general assumptions
about the $SU(6)$ symmetries of the decay amplitudes, differential
cross section and other observables \cite{j}.

Recently Zou and Riska et.al. suggested that many puzzles
surrounding baryon resonances in the CQM may be solved by extending
3-quark wave function to include multiquark configuration $qqqq\bar
q$ \cite{Li1,Li2,Li3,An,An2}. After introducing the five-quark
components, an interesting question is how the amplitudes of the
$qqq$ and $qqqq\bar q$ configurations are determined. Some authors
have discussed $\Delta(1232)$, $N^{*}(1440)$ and $N^{*}(1535)$
electromagnetic and strong decay processes which provide constrains
to the amplitudes of the $qqq$ and $qqqq\bar q$ configurations, but
only give a wide range of probability of five-quark components, one
reason of which is that these observables involve several adjustable
parameters: constituent quark mass, harmonic oscillator strengths
$\omega_{3}$ and $\omega_{5}$ of three quark and five quark
components contained in the spatial wave function.

However, the nonrelativistic axial charge operator is independent of
spatial part and  given by $ \widehat{g}_{A}
=\sum_{i}\sigma_{z}^{i}\tau_{z}^{i} $ \cite{gloz}. The axial charge
can be approximately expressed as a sum of the diagonal matrix
elements of all possible configurations,
$g_{A}$=C$P_{3}$+$\sum_{i}C_iP_{5}^{i}$ \cite{An4}, which is a sum
of axial charge $C_i$ of each single configuration multiply by
corresponding possibility $P_{i}$. Obviously axial charge is
independent of adjustable parameters mentioned above appearing in
the spatial coordinate of transition opeator. In addition, another
parameter, i.e., the relative phase factor $ \delta$ between the
$qqq$ and $qqqq\bar q$ components is not involed also. So it is easy
and reliable to obtain the relation between the probability of the
two components.

The present paper is organized as follows. In section 2, the wave
function of nucleon and nucleon resonances will be presented. In
section 3, the axial charges of the proton and nucleon resonances will be calculated in
the extended five-quark model. Conclusions and
discussions are given in the Sec.4.

\section{The five-quark wave functions of nucleon and its resonances}
\label{sec:2} For the proton and nucleon resonances, the extended
wave functions can be written as \cite{An2},

\begin{eqnarray}
|p, N^{*}(1440,1710), s_{z}\rangle&=&A_{(P)3q}|qqq\rangle
+A_{(P)5q}\nonumber\\&&\times\sum_{i}A_{i}|qqqq_{i}
\bar{q_{i}}\rangle\, ,\nonumber \\
|N^{*}(1535,1650), s_{z}\rangle&=&A_{(N^{*})3q}|qqq\rangle
+A_{(N^{*})5q}\nonumber\\&&\times\sum_{i}A_{i}|qqqq_{i}
\bar{q_{i}}\rangle\, , \label{wave}
\end{eqnarray}
where $A_{(p,N^{*})3q}$ and $A_{(p,N^{*})5q}$ are the amplitudes
factors for the 3-quark and 5-quark components, respectively. The
sum over $i$ runs over all the possible $qqqq_{i}\bar{q_{i}}$
components, i.e., $qqqu\bar{u}$, $qqqd\bar{d}$ and $qqqs\bar{s}$.

Here flavor-spin hyperfine interaction between quarks is assumed as
\cite{Hel},
\begin{equation}
C_{FS}=-\sum_{i,j}\vec\lambda^{F}_i\cdot \vec\lambda^{F}_j
\vec\sigma_i\cdot\vec\sigma_j. \label{eq1}
\end{equation}
General wave function expressions in this spin-flavor scheme for
$qqqq\bar{q}$ with positive parities are taken as
\begin{eqnarray}
&&|p,N^{*}(1440,1710),s_{z}\rangle_{5q}^{(i)}=\sum_{a,b,c,d,e}\sum_{M,m,s,S_{z}}
\nonumber\\&&C^{\frac{1}{2}s_{z}} _{JM,\frac{1}{2}s^{'}_{z}}
C^{JM}_{1m,SS_{z}} C^{[1^{4}]}_{[31]_{a}[211]_{a}}
C^{[31]_{a}}_{[31]_{b}[31]_{c}}
C^{[FS]_{c}}_{[F^{i}]_{d}[S^{i}]_{e}}\nonumber\\&\times&[211]_{C}(a)
[31]_{X,m} (b) [F^{i}]_{F}(d)
[S^{i}]_{S_{z}}(e)\bar\chi_{y,t_z}\bar\xi_{s_z}\psi({\vec{\kappa_{i}}})\,
, \ \label{pwfc}
\end{eqnarray}
where we propose that orbital state of four-quark subsystem is in
the first excited state $L=1$, whose Wely tableaux is $[31]_{X}$. $i$
is the number of the $qqqq\bar q$ configuration, as given in Table
1. $\bar\chi_{y,t_z}$ and $\bar\xi_{s_z}$ represent the wave
function of flavor and the spin space of the antiquark respectively,
and $\psi({\vec{\kappa_{i}}})$ represents the orbital symmetric
function of the momentums of the five-quark component. The first
summation involves the $S_4$ Clebsch-Gordan coefficients for the
indicated color ($[211]$), orbital space ($[31]$) and flavor ($[F]$)
and spin ($[S]$) of the $qqqq$ system, and the second one runs over
the spin indices in the standard $SU(2)$ Clebsch-Gordan coefficient.

Similarly five-quark wave function of negative parity resonance with
total angular momentum ${1\over2}$,

\begin{eqnarray}
&|N^{*}(1535,1650),t,s\rangle_{5q}^{(i)}=\sum_{a,b,c}\sum_{Y,y,T_z,t_z}\sum_{S_z,s_z}
 \nonumber\\&C^{[1^4]}_{[31]_a[211]_a} C^{[31]_a}_{[F^{(i)}]_b
[S^{(i)}]_c} [F^{(i)}]_{b,Y,T_z} [S^{(i)}]_{c,S_z}
[211;C]_a\nonumber\\&\times (Y,T,T_z,y,\bar t,t_z|1,1/2,t)
(S,S_z,1/2,s_z|1/2,s)\bar\chi_{y,t_z}\bar\xi_{s_z}\nonumber\\&\times\varphi_{[5]}\,
, \label{wfc}
\end{eqnarray}
where we assume that orbital state of four-quark subsystem is in the
ground state L=0, whose Wely tableaux is $[4]_{X}$. So orbital wave
function $\varphi_{[5]}$ is completely symmetric.

These wave functions are given in explicit form in Ref. \cite{An2}.
Now,we can order these configurations in the terms of increasing
matrix element $\langle C_{FS} \rangle$ of chiral hyperfine
interaction (2) \cite{Hel}. The $qqqq\bar q$ configurations having
appropriate quantum numbers in the proton, $N^{*}(1440)$ and
$N^{*}(1710)$, and the corresponding axial charge coefficient $C_i$
are listed in Table \ref{co1} and for the $N(1650)$ and $N(1535)$,
the results can be found in Ref.\cite{An4}.

\begin{table}
\caption{The $qqqq\bar q$ configurations having appropriate quantum
numbers in the proton, $N^{*}(1440)$ and $N^{*}(1710)$, and the
corresponding coefficients $C_i$ in the axial charge expression.
\label{co1}}
%\begin{ruledtabular}
\begin{tabular}{|c|c|c|c|c|}
\hline configuration & flavor-spin&$\langle C_{FS} \rangle$&$C_i$
\\
\hline 1& $ [4]_{FS}[22]_F [22]_S $ & $-28$& $-2/9$ \\
2& $ [4]_{FS}[31]_F [31]_S $ & $-64/3$ &$-4/15(J_{q^{4}}=0)$   \\
3& $ [4]_{FS}[31]_F [31]_S $ & $-64/3$ &$28/45(J_{q^{4}}=1)$   \\
4& $ [31]_{FS}[211]_F  [22]_S $ &$-16$ &$0$   \\
5& $ [31]_{FS}[211]_F  [31]_S $ &$-40/3$&  $0(J_{q^{4}}=0)$ \\
6& $ [31]_{FS}[211]_F  [31]_S $ &$-40/3$&  $4/9(J_{q^{4}}=1)$ \\
7& $ [31]_{FS}[22]_F  [31]_S $ & $-28/3$   &$+17/18$\\
\hline
\end{tabular}
%\end{ruledtabular}
\end{table}

\section{The axial charges of the proton and its resonances with the extended five-quark wave function}
%\label{sec:3}

It is straightforward to calculate the diagonal charges in the
$SU(6) \bigotimes O(3)$ symmetry scheme \cite{gloz}. Then we
consider the mixings between nonstrange baryons due to symmetry
breaking and find that the prediction is not well improved for
proton. The mixing coefficients for proton and $N^{*}(1440)$ are
taken from Ref.\cite{j}. Without configuration mixings axial charges
is given in the first row and with mixings axial charges given in
the third row in Table \ref{co3}. In the last row, the values are
obtained in the extended quark model.

\begin{table}[ht!]
\caption{Axial charges of the proton, $N^{*}(1440)$ and
$N^{*}(1710)$ in the both unmixing and mixing cases.\label{co3}}

\begin{tabular}{|c|c|c|c|}
\hline  &$g_{A}(proton)$ & $g_{A}(N(1440))$& $g_{A}(N(1710))$

\\
\hline  Unmixing & $5/3$ &$5/3$&$1/3$\\
Mixing  & ${1.60}$&${1.67}$&$-$\\
ECQM& $1.25\sim1.46$&$\sim1$&$-$\\

\hline
\end{tabular}
\end{table}

\begin{table*}
    \caption{Helicity ampllitudes  in units of $GeV^{-1/2}$.
$10\%\sim20\%$ five-quark components in the proton is
assumed.\label{co4}} \centering
\begin{tabular}{cccccc}
\hline
   $A(N^{*}\rightarrow N\gamma)$    & $A_{1/2}$  &                  &    $A_{3/2}$
  &   & \\
    &           exp.            &  $10\%\sim20\%$    & exp.
 &   $10\%\sim20\%$&\\
\hline
 $N^{*}(1440)$ & $-0.065\pm0.004$& $-0.045\sim-0.056$ & $-$ & $-$ &\\

 $\Delta(1232)$ &$-0.135\pm0.006$& $-0.093\sim-0.091$&$-0.255\pm0.008$& $-0.167\sim-0.171$ &\\

 \hline
\end{tabular}

\end{table*}

%%%%%%%%%%%%%%%%%%%%%%%%%%%%%%%%%%%%%%%%%%%%%%%%%%%%%%

In the extended constituent quark model (ECQM), with the values in
Table 1, the diagonal axial charge reads
\begin{eqnarray}
g_A (N(938,1440))&=& {5\over 3}P_3 - {2\over 9} P_5^{(1)}-{4\over
15} P_5^{(2)J_{q^{4}}=0}+{28\over 45} \nonumber\\&&\times
P_5^{(3)J_{q^{4}}=1} +{4\over 9}P_5^{(5)J_{q^{4}}=1}\, , \label{eq3}
\end{eqnarray}
where $P_3$ represents the probability for the $qqq$ configuration,
while $P_5^{(i)}$ is the probability for i-th $qqqq\bar q$
configuration in Table \ref{co1}.
%%%%%%%%%%%%%%%%%%%%%%%%%%%%%%%%%%%%%%%%%%%%%%%%%%%%%%

The energy of the configuration
with the spin-flavor symmetry $ [4]_{FS}[22]_F [22]_S $ is about
140-200 $MeV$ lower than others within the hyperfine interaction
model of Eq.(\ref{eq1}) and simultaneously can give the
experimentally observed positive strangeness magnetic moment and
strange electric radii \cite{zou1}, and negative strange electric
form factor \cite{ris}. If only the first two terms in
Eq.(\ref{eq3}) are taken into consideration, $g_A (N(938))$ would be
equal to the experimental value 1.26 with $P_5^{(1)}\sim 20\%$.
Electromagnetic and strong decays of $\Delta(1232)$, $N^{*}(1440)$
\cite{Li1,Li2,Li3} and $N^{*}(1535)$ \cite{An2} also put strong
restrictions on five quark proportion in the proton and leads to the
probability of $qqqq\bar{q}$ components in the proton about
$10\%\sim20\%$. This proportion is inserted into the expression
(\ref{eq3}) and the obtained numerical value for $g_A (N(938))$
falls in the range 1.25 to 1.45 which brackets 1.26. In the same
time, the change of helicity amplitudes for $\Delta(1232)$ and
$N^{*}(1440)$ is by only a few percent, as given in table \ref{co4},
if the probability of five-quark components in the proton is assumed
to fall in the range $10\%\sim20\%$.

For the $N^{*}(1440)$, by explicit introduction of about $30\%$ five
quark component it becomes possible to improve the helicity
amplitude and strong decay width of $N^{*}(1440)$. If only the
configuration $[4]_{FS}[22]_F [22]_S $ is taken into account,
inserting $P_5^{(1)}\sim 30\%$ into Eq.(\ref{eq3}) will give axial
charge of $N^{*}(1440)$ about unity. However, chiral restoration
scheme predicts that the chiral partner of\\ $N^{*}(1535)$, Roper
state, also has the small axial charge, i.e., $\sim0$.

The next-to-lowest-energy $qqqq$ configuration is $[4]_{FS}$ $[31]_F$ $
[31]_S $. The $J_{q^{4}}=0$ and $ J_{q^{4}}=1$ $qqqq$ states of
$N^{*}(1440)$ are assumed to have the equal proportion, i.e.,
\begin{equation}
P_5^{(2)J_{q^{4}}=0}=P_5^{(3)J_{q^{4}}=1}={1\over2}P_5^{(2)} ,
\label{gax}
\end{equation}
where $P_5^{2}$ is the probability of spin-flavor symmetry
$[4]_{FS}$ $[31]_F $ $[31]_S $ of $qqqq$ subsystem in $N^{*}(1440)$. With
the same total proportion ($\sim 30\%$) for five-quark component in
$N^{*}(1440)$ as before, but the probability of $[4]_{FS}[22]_F
[22]_S $ and $[4]_{FS}[31]_F [31]_S $ taken as $80\%$ and $20\%$,
respectively, the final $g_A (N(1440))$ is obtained as 1.04. In
comparison to axial charge with only the configuration
$[4]_{FS}[22]_F [22]_S $ considered, the new configuration does not
lead to any obvious change.

As for $N^{*}(1535)$, lattice QCD results show that axial charge
takes quite small value. It is possible for $N^{*}(1535)$ to have
the very small or possibly vanishing axial charge after considering
the sea quarks configurations \cite{An4}. In the CQM, after taking
into account mixing angle between $N^{*}(1535)$ and $N^{*}(1650)$,
the predictions on both axial charges have large deviations from
lattice results. In the calculation, standard mixing angle
$\theta_{S}$ from OGE in the lowest mass negative parity nucleons is
$-32^{\textordmasculine }$. We take into consideration the
experimental error of $10^{\textordmasculine }$ and calculate the
axial charges, as given in the table \ref{co5}, of $g_{A}(N(1535))$
and $g_{A}(N(1650))$ at mxing angles, $-22^{\textordmasculine }$ and
$-42^{\textordmasculine }$, respectively. Without configuration
mixings axial charges is given in the first column and with mixings
axial charges given in the other three columns. Lattice results of
axial charge is given in the sixth column \cite{ta}. In the last
column, the values are axial charge obtained in the extended quark
model.

Physical resonances can be expressed as
\begin{eqnarray}
|N^{*}(1535)\rangle= \cos\theta _{S}|^{2}P_M\rangle - \sin\theta
_{S}|^{4}P_M\rangle, \nonumber \\
 |N^{*}(1650)\rangle=
\sin\theta_{S}|^{2}P_M\rangle +\cos\theta _{S}|^{4}P_M\rangle .
\label{gax}
\end{eqnarray}
Here the $\theta_{S}$ is mixing angle, which is defined as Isgur and
Karl \cite{is}. The axial charges $N^{*}(1535)$ and $N^{*}(1650)$
are defined by
\begin{eqnarray}
g_{A}(N(1535))&=&
\cos^{2}\theta_{S}\langle^{2}P_M|\sum_{i}\sigma_{z}^{i}\tau_{z}^{i}|^{2}P_M\rangle
+\sin^{2}\theta_{S}
\nonumber\\&&\times\langle^{4}P_M|\sum_{i}\sigma_{z}^{i}\tau_{z}^{i}|^{4}P_M\rangle-
2\cos\theta_{S}
\nonumber\\&&\times\sin\theta_{S}\langle^{2}P_M|\sum_{i}\sigma_{z}^{i}\tau_{z}^{i}|^{4}P_M\rangle,
\nonumber\\
g_{A}(N(1650))&=&
\sin^{2}\theta_{S}\langle^{2}P_M|\sum_{i}\sigma_{z}^{i}\tau_{z}^{i}|^{2}P_M\rangle
+\cos^{2}\theta_{S}\nonumber\\&&\times\langle^{4}P_M|\sum_{i}\sigma_{z}^{i}\tau_{z}^{i}|^{4}P_M\rangle
+
2\cos\theta_{S}\nonumber\\&&\times\sin\theta_{S}\langle^{2}P_M|\sum_{i}\sigma_{z}^{i}\tau_{z}^{i}|^{4}P_M\rangle,
\label{gax}
\end{eqnarray}

The sum of both axial charges is a variable independent on the
mixing angle, i.e., $g_{A}(N(1535))$+$g_{A}(N(1650))$ = $\frac{4}{9}$
exactly is a constant. The lattice result is about 0.55. We see that
it is impossible to improve the prediction of axial charges in the
CQM by adding the effects of the $SU(6) \bigotimes O(3)$ symmetry
breaking.

\begin{table*}
\begin{center}
 \caption{Axial charges  of the two lowest-energy negative
parity resonances.  \label{co5}}
\begin{tabular}{|c|c|c|c|c|c|c|}
\hline  & Unmixing &$ \theta _{S}=-22^{\textordmasculine }$&$ \theta
_{S}=-32^{\textordmasculine }$& $\theta _{S}=-42^{\textordmasculine
}$&Lattice&ECQM
\\
\hline  $g_{A}(N(1535))$ & $-1/9$ &0.29&0.47&0.63&$\sim0$&$-0.05\sim+0.06$\\
  $g_{A}(N(1650))$ & $5/9$&0.15&-0.03&-0.19&$\sim0.55$&$\sim0.5$\\
\hline
\end{tabular}
\end{center}

\end{table*}
%%%%%%%%%%%%%%%%%%%%%%%%%%%%%%%%%%%%%%%%%%%%%%%%%%%%%%%

The strong couplings of $N^{*}(1535)N\phi$, $N^{*}(1535)N\eta$ and
$N^{*}(1535)K\Lambda$, as predicted in Refs. \cite{xie,zou2}, show
that there may be large strangeness components in $N^{*}(1535)$
resonance. In addition, it is proved to appear possible to reach the
value 0 for axial charge in the extended quark model \cite{An4}. And
after about $ 45\%$ $qqqq\bar{q}$ components taken
into account, the helicity amplitudes for the electromagnetic
transition $\gamma^{*} N\rightarrow N^{*}(1535)$ in the extended
quark model have satisfactory description \cite{An2}.

In the case of $N^{*}(1650)$, theoretical predictions for $\pi$ and
$\eta$ decay withds of the $N^{*}(1650)$ in the traditional quark
model are unsatisfactory \cite{pl}. A more realistic resonance wave
function beyond pure three-quark bound state, including five-quark
component, may improve the theoretical expectations. Like
$N^{*}(1535)$ \cite{An4}, we have the following expression of axial
charge,
\begin{equation}
g_A (N(1650))= {5\over 9}P_3+{5\over 6}P_5^{(2)} - {1\over 9}
P_5^{(3)}-{4\over 15} P_5^{(4)}-{17\over 18}P_5^{(5)} \, .
\label{gax}
\end{equation}

If large range $10\%\sim30\%$ of proportion of five-quark component
in $N^{*}(1650)$ is assumed, the numerical value for its axial
charge falls in the range 0.39 to 0.54, which includes the lattice
value $\sim0.5$.

\section{Conclusions}
We conclude that axial charge of proton in the traditional quark
model does not have the substantial improvement even though the
breakdown of the $SU(6) \bigotimes O(3)$ symmetry is taken into
account. After taking relativistic effects into account, for
massless relativistic quarks in the MIT bag model, one obtains much
smaller value $g_A (N(938))$=1.09, which is also in disagreement
with experiment value. But the empirical data for the axial charge
of the proton can be well described qualitatively in the extended
constituent quark model including the five-quark component, and the
probabilities of the five-quark component ($\sim 20\%$)  we obtain
in the proton are consistent with the previous results \cite{An}. In
addition, we predict a unity axial charge for $N^{*}(1440)$, if the
contributions of $\sim 30\%$ five-quark components are considered.
In the future more wealth lattice data of axial charge will put
constrain on proportion of sea-quark components in Roper state and
other nucleon resonances.

For $N^{*}(1650)$ it is unknown whether or not it has sizable five
quark components. Its properties might be explored in the
photoproduction of the kaon meson through analyzing the wealthy
polarized and unpolarized observables data. And in the strong
interaction process $pp \rightarrow pK^{+}\Lambda$, new measurements
at excess energy 171 $MeV$ also clearly show the contribution of
$S_{11}(1650)$ baryon excitation \cite{as}. In order to obtain
$S_{11}$ properties we could analyze the more rich data at CSR or
COSY in the future.

\section{Acknowledgments}

S. G. Yuan acknowledges the hospitality of the large scientific
facility theoretical center during the course of the work.

\end{document}